\documentclass[11pt]{article}
\usepackage{geometry} 
\usepackage{graphicx}
\usepackage{amssymb}
\usepackage{epstopdf}
\usepackage{amsmath}
\usepackage{fixmath}
\usepackage{MnSymbol}
\usepackage{stmaryrd}
\usepackage{amsfonts}
\usepackage{array}
\usepackage{multirow}
\usepackage[affil-it]{authblk}

\DeclareMathAlphabet{\mathpzc}{OT1}{pzc}{m}{it}

\title{Hidden Equilibration Driven Losses in Whitecapping}
\author[1]{Clifford E Chafin}
\affil[1]{Physics Dept., North Carolina State University, Raleigh NC, 27695}

\begin{document}
\maketitle
\begin{abstract}
The role of whitecapping losses of waves is investigated in a simple model based on conservation laws.  It is shown that, for Airy waves, at least as much energy is lost in gradual reequilibration as is lost in the whitecapping events themselves.  This model is based on the the notion that the waves and losses are small enough that some narrow spectrum of frequencies reappears over time.  
\end{abstract}

The mechanisms of growth of wind blown waves are still somewhat in dispute.  The Miles/Phillips mechanism \cite{Miles,Phillips, Lighthill} incorporating wave sheltering is the most popular theory.  Recently, the author established some bounds on the size of growth from sheltering versus ``crest deposition'' whereby surface shear on the windward side drives mass upwards to increase wave height \cite{Chafin-cd}.  Angular momentum is a subtle to track quantity in extended continua particularly when it comes to consideration of when it is locally conserved in waves versus in shear flow or end-of-packet contributions.  Waves can undergo interaction and tend to drift towards a more narrow distribution of frequencies.  Angular momentum can then place constraints on the efficiency of wave growth.  

Waves can damp by viscous losses into surface shear and destructive turbulent effects of breaking where angular momentum gets exchanged with the ocean bottom and vorticity is rapidly pumped into the surface flow.  
In this article, we consider an analogous situation for the case of wave whitecapping.  Specifically, we are interested in the case of removal of water, generally by wind, from the tops of waves in a sea that has no driving forces and the interactions drive the waves towards near monochromaticity.  We assume that the removed water is dropped back into the sea in such a way that it makes no contribution to future wave motion.  The Benjamin-Feir instability gives a side band broadening and results in eventual formation of a series of pulses \cite{BF}.  We assume that whatever waves we are dealing with, this does not occur or does not do so on any time scale of concern so the monochromatic assumption has  value here as an approximation. 

 Our calculations are done using Airy waves.  We conclude that waves that lose energy and momentum to whitecapping must lose at least as much momentum in such an equilibration process.  This is shown to be entirely due to conservation of linear and angular momentum and the tendency of waves to gravitate towards a monochromatic distribution without continued driving.  
 
The linearized deep water solutions to the Navier-Stokes equations give surface waves described by the basis of pairs
\begin{align}
\phi_{k}(x,z)&=e^{kz}\cos(kx)\\
\eta_{k}(x,t)&=\frac{k}{\omega}\phi_{k}(x,0)\sin(\omega t)\\
\phi_{k}^{\dagger}(x,z)&=e^{kz}\sin(kx)\\
\eta_{k}^{\dagger}(x,t)&=-\frac{k}{\omega}\phi_{k}(x,0)\cos(\omega t)
\end{align}
for the surface deformation $\eta$ and velocity potential $\phi$.  These can be linearly combined to give purely progressive waves.  The free surface advance of these allow for a small mass flux called Stokes drift.  The conserved quantities associated with these waves are tabulated in tab.\!~\ref{tab} \cite{Chafin-rogue}.  The mass elevation is a necessary consequence of long range pressure fields of packets and maintaining Stokes drift with the support of the packet.  

\begin{figure}[!ht]
   \centering
   \includegraphics[width=3in,trim=0mm 30mm 0mm 0mm,clip]{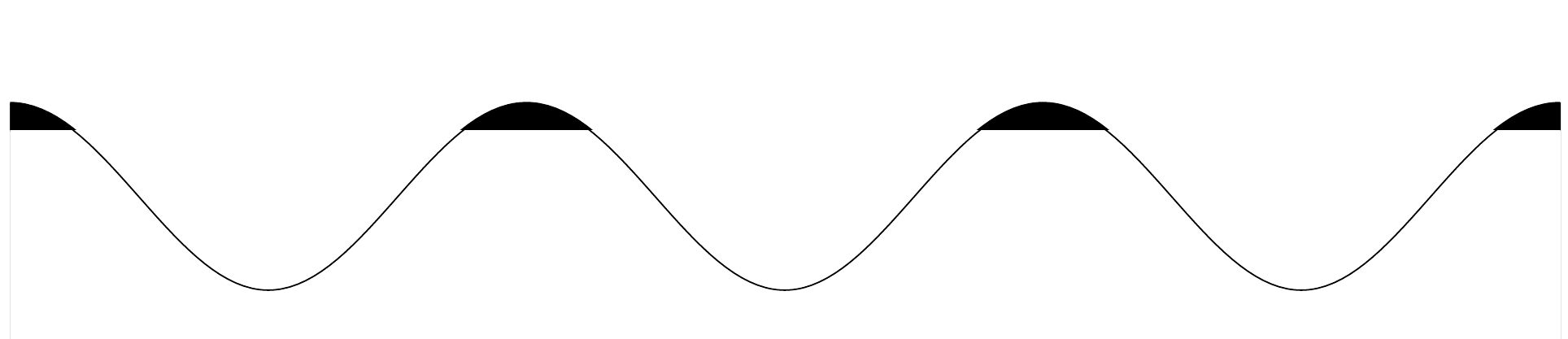} 
   \caption{The removed layer of fluid from idealized white capping and surface spray.}
   \label{chopped}
\end{figure}

Now let us consider the behavior of a progressive wave and introduce the (idealized) effect of wave breaking and surface spray.  This is illustrated in Fig.~\ref{chopped} where we model the new wave as the old one with a ``trimmed'' off layer.  The removed layer contains energy and momenta of the wave (and a contribution from the acceleration from the wind).  In our idealized limit, wind driven shear and spray removes it from wave motion but redeposits it as a shear flow and heating of the surface layer.  To understand its effect we analyze the energy and momenta removed from the wave assuming the process is slow enough to be ``adiabatic.''

The conserved quantities of Airy waves are expressed in Table~\ref{tab}.
\begin{table}
\begin{center}
\vspace{0.5cm}
\begin{tabular}{|l|l|l|l|}  \hline $\mathcal{E} $& $\mathit{m}$&$\mathpzc{p}$ & $\mathcal{L}$ \\  \hline \multirow{2}{*}{$\frac{1}{2}\rho g a^{2}$}& \multirow{2}{*}{$\rho a^{2}  k$}  & \multirow{2}{*}{$\frac{1}{2}\rho a^{2} \omega $} &\multirow{2}{*}{$-\frac{1}{4}\rho g \frac{ a^{2}}{\omega}$}  \\ &  &  &\\ \hline \end{tabular} 
\vspace{0.5cm}
\end{center}
\caption{Conserved depth-integrated and time averaged quantities.}\label{tab}
\end{table}
The mass change is chosen to give the elevation a packet needs to carry Stokes drift.  It is result of long range pressure changes as in Chafin \cite{Chafin-rogue}.  
The crests of the waves are modeled as parabolas of the form $a(1-\frac{1}{2}k^{2}x^{2})$.  We remove a layer of fluid of thickness $\delta=a\epsilon^{2}$.  The mass, energy and momenta contained in these layers per net unit length is


\begin{align}
d\mathit{m}&=\frac{2}{3}\frac{\sqrt{2}}{\pi}\rho a\epsilon^{3}\\ 
d\mathcal{E}&=\frac{2}{3}\frac{\sqrt{2}}{\pi}\rho g a^{2}\epsilon^{3}=(a g) d\mathit{m}\\
d\mathpzc{p}&=\frac{2}{3}\frac{\sqrt{2}}{\pi}\rho a^{2}\omega\epsilon^{3}=(a \omega) d\mathit{m}\\
d\mathcal{L}&=-\frac{4}{15}\frac{\sqrt{2}}{\pi}\rho a^{2}\frac{\omega}{k}\epsilon^{5}=-\frac{2}{5} \left(\frac{\omega }{k}\delta\right)d\mathit{m}
\end{align}
where the contribution to the energy is all PE to this order.  We see that $\frac{dp}{dE}=\frac{\omega}{g}=v_{ph}^{-1}$ and $\frac{k}{\omega} = \frac{p}{E}$.  The angular momentum however does not change to the same order as the energy and momentum so it can be ignored.

We have a situation where energy and momentum are being drawn off of the wave and the wave is simultaneously evolving.  Constancy of angular momentum enforces the relation
\begin{align}
\dot{a}=\frac{1}{2}a\frac{\dot{\omega}}{\omega}
\end{align}
Assume we are removing the tops of the waves at a mass loss rate rate $\dot{m}$ and all the remaining momentum remains in the waves.  The momentum lost in this process from the waves is
\begin{align}
\dot{\mathpzc{p}}=2\rho a \omega \dot{a}=a\omega \dot{m}
\end{align}
so that $\dot{a}=\frac{\dot{m}}{2\rho}$.  Computing the energy lost from the wave from this change of height we have 
\begin{align}
\dot{\mathcal{E}}=\frac{1}{2}g a \dot{m}
\end{align}
but we see that this is only half of the energy lost from the mass actually removed by the crest trimming $\dot{\mathcal{E}}=g a \dot{m}$.  It is possible that the actual energy loss could be greater than this but certainly not less.  As a consequence the efficiency of energy removal from the whitecapping is, at most, 50\%.  The rest must dissipate into surface flows in the process of equilibration.  

Nonlinear waves, like Stokes waves, have a higher ratio of momentum to energy in the flow near the crests while that of the angular momentum is reduced so that such an effect is probably reduced for such steeper waves where whitecapping is most prevalent.  In such a case, the shear driven crest deposition is also generally stronger so there will be competing effects damping and driving the waves.  The number of complications to wave growth and evolution is large but this one does not seem to have been addressed from the standpoint of conservation laws.  The elevation changes from long range pressure fields and microbreaking are generally orders of magnitude smaller than the wavelengths.  These complicate numerical results at a scale where the long time evolution is often determined.  In a subject dominated by perturbation theory and numerical simulation, being able to place a strong bound on behavior based on conserved quantities should be considered a uniquely potent result.

\end{document}